# MAGNet: A Multi-Scale Attention-Guided Graph Fusion Network for DRC Violation Detection

Weihan Lu and Hong Cai Chen, *IEEE Member*

*Abstract*—Design rule checking (DRC) is of great significance for cost reduction and design efficiency improvement in integrated circuit (IC) designs. Machine-learning-based DRC has become an important approach in computer-aided design (CAD). In this paper, we propose MAGNet, a hybrid deep learning model that integrates an improved U-Net with a graph neural network for DRC violation prediction. The U-Net backbone is enhanced with a Dynamic Attention Module (DAM) and a Multi-Scale Convolution Module (MSCM) to strengthen its capability in extracting fine-grained and multi-scale spatial features. In parallel, we construct a pixel-aligned graph structure based on chip layout tiles, and apply a specialized GNN to model the topological relationships among pins. During graph construction, a graph-to-grid mapping is generated to align GNN features with the layout image. In addition, a label amplification strategy is adopted during training to enhance the model's sensitivity to sparse violation patterns. Overall, MAGNet effectively combines spatial, semantic, and structural information, achieving improved prediction accuracy and reduced false positive rates in DRC hotspot detection. Subsequently, through incremental training, we achieve a more sensitive discrimination ability for hotspots. The results demonstrate that, in comparison with ibUnet, RouteNet, and J-Net, MAGNet significantly outperforms these models, achieving substantial improvements in overall performance.

## I. Introduction

Integrated circuits (IC) are playing an increasingly important role in modern electronic designs [1]. As the growing complexity of ICs, design rule checking (DRC) becomes more important. DRC is performed in the backend stage of chip design. The main purpose is to ensure that the layout meets the manufacturing process requirements and to avoid manufacturing defects. Traditional DRC relies on rule-based verification tools, which, although precise, are computationally intensive. For full-chip-level complex designs, the verification process may last for several hours or even days.

In recent years, using machine learning (ML) for DRC has become an important method [2] in computer-aided design (CAD). A prominent direction involves modeling layout features as images and applying convolutional neural networks (CNNs) to detect patterns and identify rule violations [3–5].

This work is funded by Fundamental Research Funds for the Central Universities (No. 3208002309A2). (Corresponding author is Hong Cai Chen)

Weihan Lu and Hong Cai Chen are with School of Automation, Southeast University, Nanjing 210096, China. (e-mail: 213221757@seu.edu.cn , chenhc@seu.edu.cn)

Among these, the U-Net architecture has proven especially effective due to its ability to capture fine-grained spatial structures through its encoder-decoder design [6–9]. A major advancement in this area was the release of CircuitNet in 2024 [14]. This dataset provides nine critical layout features, enabling more standardized and reproducible benchmarks for DRC learning tasks. U-Net-based models built upon CircuitNet have since become a dominant baseline due to their simplicity and effectiveness.

In addition to CNNs, alternative approaches such as graph neural networks (GNNs) have been introduced to capture complex rule interactions and layout connectivity [11–13]. These diverse strategies illustrate the increasing depth and breadth of ML-driven DRC research.

Furthermore, A study has analyzed DRC at the pre-route stage [23], which has significantly increased the number of resolved DRC violations. A study [24] proposed a Transformer-based clustering algorithm for standard cell design, which generated more DRC/LVS-clean layouts and achieved a speedup over previous techniques. Beyond modeling layout features as images, some researchers have employed point cloud representations, viewing circuit elements as spatial point sets to extract geometric and topological features [10, 26]. This highlights the rising influence of modeling in physical design automation.

However, despite these advances, many recent works lack comprehensive evaluation using critical metrics such as False Positive Rate (FPR), True Positive Rate (TPR), and F1-score, limiting the interpretability of performance claims. Additionally, the abstraction introduced by image-based modeling—as in CircuitNet—while beneficial for model efficiency, may overlook key physical routing semantics, such as layer-specific interactions or metal congestion. This simplification can compromise the detection accuracy for rule violations that are sensitive to detailed routing structures.

These limitations underscore the need for more holistic DRC models that combine high-level abstraction with low-level physical awareness. As effective DRC requires the joint consideration of pin accessibility and routing congestion, we propose a hybrid modeling approach to achieve more comprehensive feature extraction.

Unlike previous models that rely solely on CNN-based image segmentation or use a single large-scale graph. In this paper, we propose MAGNet, a novel DRC detection framework that combines U-Net and GNN architectures to jointly capture spatial and topological features of the layout. Meanwhile, a spatial mapping mechanism is employed to guide the integration of features from different branches, enabling more effective fusion and improved feature extraction. MAGNet introduces a tile-based hybrid approach that combines localized graph reasoning with pixel-level feature learning.

The main contributions of the proposed MAGNet can be summarized as follows:
1) We propose a novel pixel-based graph construction strategy which partitions the layout into fixed-size subregions and builds a localized graph for each tile. This localized modeling captures fine-grained pin connectivity and obstacle features while avoiding the scalability issues of global layout graphs.
2) We design an enhanced U-Net variant, named MD-Unet, which incorporates a Multi-Scale Convolution Module (MSCM) and a Dynamic Attention Mechanism (DAM) to improve layout-level DRC prediction. These enhancements improve the network's ability to extract robust and discriminative features from multi-channel layout inputs, leading to more accurate prediction of DRC violation distributions.
3) We design a cross-modal feature fusion strategy that combines the semantic features from MD-Unet with the topological cues from GNN during the prediction stage. A tile-based map, generated during graph construction, ensures spatial consistency throughout the fusion process, thereby improving hotspot localization accuracy and reducing the false positive rate.

Despite the increased architectural complexity, MAGNet maintains computational efficiency by leveraging the independence of pixel-based graph, which enables parallel GNN processing across tiles. This tile-wise design ensures scalability and tractability, even on large layouts. The following parts of this paper are composed as follows: Section II introduces key components of MAGNet including the basic knowledge of U-NET and GNN. Section III provides a detailed introduction to the model construction. Section IV introduces our training processes and the experimental results of the model. Finally, Section V summarizes this paper.

## II. FUNDAMENTAL KNOWLEDGE

U-Net excels at capturing local spatial patterns from dense layout features, while GNN is effective at modeling the topological relationships between circuit components. The following sections provide a brief overview of the fundamentals of U-Net and GNN.

### A. U-Net Model

U-Net was initially applied to medical image segmentation and has been proved to show excellent abilities in image segmentation and image processing [15, 28]. The main components of U-Net are the encoder, decoder, and skip connections. The encoder mainly consists of many convolutional layers and pooling layers, which reduce the input image and increase the dimension through the encoder's action. The decoder is symmetric with the encoder and obtains the target image through the reverse decoding process. The skip connection part is the main part of machine learning. The connection of corresponding layers enables U-NET to learn parameters and achieve the target function [27].

$$Feature^{DRC} \in \{0,1\}^{w*h*f} \to Label^{DRC} \in \{0,1\}^{w*h*1} \quad (1)$$

Equation (1) represents the relationship between the input and the output. The specific data structure of U-Net is consistent with that in Fig. 1.

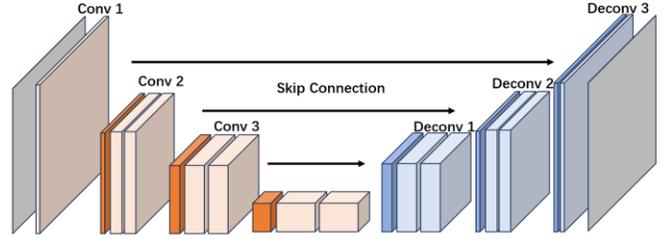

Fig. 1. The basic structure of U–Net.

### B. Graph Neural Networks (GNNs)

GNNs are a class of deep learning models specifically designed to operate on graph-structured data. Many problems in electronic design naturally exhibit graph-like structures—such as netlists, routing paths, and layout connectivity—making GNNs well-suited for a wide range of tasks in this domain [32].

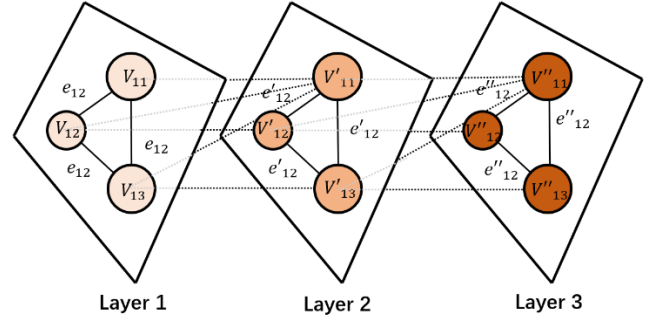

Fig.2. The basic structure of GNN.

Formally, a graph is defined as $G = (V, E)$, where $v$ is the set of vertices (nodes) and $e$ is the set of edges. Each node $v_i \in V$ is associated with a feature vector, and the edges are represented by an adjacency matrix $A \in \{0,1\}^{n*n}$, where $n = |V|$. The node features can be organized into a node matrix $X \in R^{n*d}$, where $d$ is the feature dimension.

A GNN processes the graph through a series of graph convolutional layers, where the key operations are message passing and neighborhood aggregation. In each layer, every node updates its embedding by aggregating features from its neighboring nodes, typically through a learnable function. This process allows nodes to iteratively capture both local and higher-order structural information.

## III. DESIGN OF THE MAGNET

The proposed MAGNet integrates a novel multi-scale & attention-enhanced U-Net and a tile-based graph with a supporting GNN through map-guided feature fusion. The architectural details of each component are presented in the subsequent sections.

### A. MD-Unet: Multi-Scale & Attention-Enhanced

MD-Unet is constructed based on the classical U-Net architecture, with modifications to enhance its capability in detecting layout hotspots.

- **Multi-Scale Convolution Module (MSCM)**

In traditional U-Net architectures, convolutional layers typically use fixed-size kernels, which limits the network's ability to simultaneously capture features at different spatial resolutions. This poses a challenge in DRC violation detection, where layout patterns can vary significantly in scale—for example, from narrow wire segments to large macro blocks. Relying solely on a single kernel size may cause the model to overlook either fine-grained violations or broader structural context.

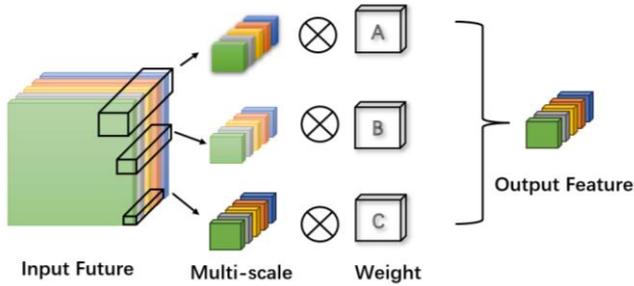

Fig. 3 The schematic of MSCM.

To address this limitation, we propose the use of MSCMs to replace the standard convolutional layers in the U-Net backbone. Each MSCM applies multiple convolution kernels of different sizes—specifically [3×3], [5×5], and [7×7]—in parallel to the same input feature map. [18,19] The multi-scale convolution method has been proved to perform well in multi-dimensional input problems.

$$F_i = Conv_{ki}(F) \quad k_i = 1,2,3$$

$$F^{out} = \sum w_i \cdot F_i$$

(2)

Equation (2) shows how the MSCMs works. The outputs are then aggregated through a weighted summation to form a unified feature representation. This design allows the network to concurrently capture small-scale details (e.g., pin spacing, thin wires) and large-scale layout structures (e.g., metal regions, macro boundaries). A schematic of the MSCM structure is shown in Fig. 3.

- **Dynamic Attention Mechanisms (DAMs)**

To further improve model expressiveness, dynamic attention mechanisms (DAMs) are introduced. The dynamic attention mechanism has been proved to be suitable for problems with multi-dimensional inputs [16, 17, 25]. By enhancing the weights of important dimensions and suppressing the weights of unimportant dimensions, the attention effect is achieved. When using U-NET to solve DRC problems, since the input is a 9-dimensional image and the dimension will be further increased during the encoding and decoding processes, we can use the dynamic attention mechanism to solve DRC problems. The dynamic attention includes spatial attention and channel attention.

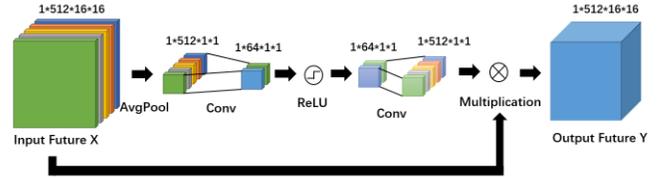

Fig.4. The structure of channel attention.

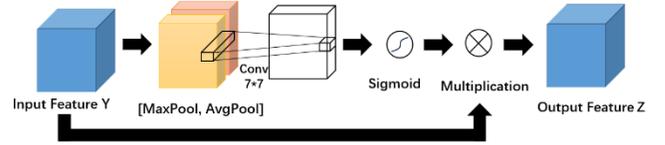

Fig. 5. The structure of spatial attention.

$$Z_c = \frac{1}{H*W} \sum_{i=1}^{H} \sum_{j=1}^{W} F_c(i,j)$$

$$a_c = \sigma(w_2(w_1 * Z_c + b_1) + b_2)$$

(3)

$$w_1 \in R^{\frac{C}{r}*c}, w_2 \in R^{c*\frac{C}{r}}, b_1 \in R^{\frac{C}{r}}, b_2 \in R^C$$

$$F^{out} = a_c \otimes F$$

Equation (3) represents the function of channel attention. σ is the Sigmoid function, $Z_c$ represents the average pooling of a certain channel, Ac represents the attention weight, $F^{out}$ represents the feature map after adding the attention, F represents the feature map without adding the attention, and ⊗ represents the element-wise multiplication by channel. We can know the structure of channel attention in Fig. 4.

$$F_c = \sigma\left(cove([F_{max}; F_{avg}])\right) \quad F_{max}, F_{avg} \in R^{1*H*W}$$

$$F^{out} = F \otimes F_c$$

(4)

Equation (4) represents the function of spatial attention. σ is the Sigmoid function, $F_{avg}$ represents the average pooling feature map, $F_{max}$ represents the maximum pooling feature map, $F^{out}$ represents the output feature map, F represents the input feature map, and $\otimes$ represents the element-wise multiplication. We can know the structure of spatial attention in Fig. 5.

Attention is only applied to the feature map at the final encoder layer, which has the size [16×16×512]. Since this layer has the highest number of channels, the application of channel attention at this stage is most effective.

The combination of MSCMs and DAMs forms the enhanced U-Net, denoted as MD-Unet. A simplified schematic of MD-Unet is shown in Fig. 6. The network adopts a 4-layer encoder and a 5-layer decoder structure. The input to MD-Unet is a 3D tensor of shape [256×256×9], representing 9 feature channels of the layout image. The output is a predicted hotspot probability map of size [256×256×1].

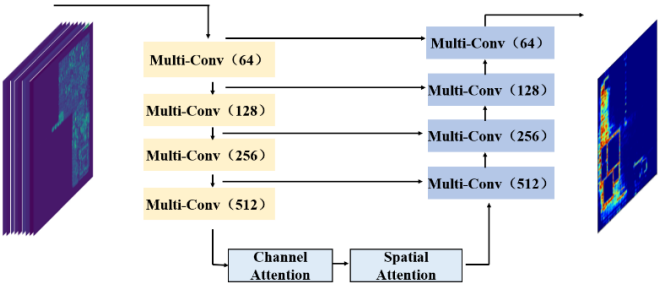

Fig. 6. The overall structure of MD-Unet.

*B. A tile-based graph & Supporting GNN*

While MD-Unet captures pixel-level spatial features, it lacks awareness of topological structures critical for layout rule checking. To compensate for this, we introduce GNN, a graph neural network operating on pixel-based graph. To adapt to the complexity of graphs, multi-layer graph is proved to be improved [29, 30].

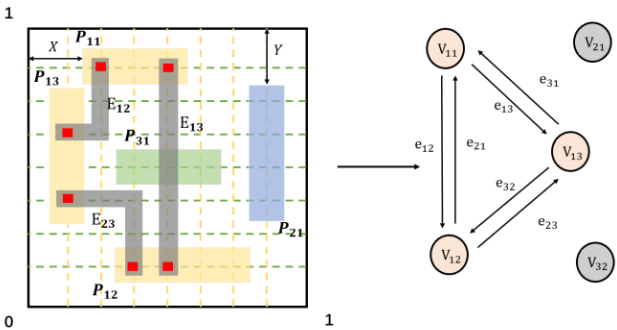

Fig. 7 The example of tile-based graph.

*A tile-based graph* is a special graph designed to represent the connectivity of chip pins. We divide the chip into multiple grids, each with a fixed size of 256×256, and each grid contains information about the chip wiring within it. It is often proposed that there are four factors influencing pin connectivity: 1) The shape of the pins. 2) The connection mode of pins in the network. 3) The mutual influence between pins. 4) The number of pins in the network. To reflect the features, PGNN [11] proposed two models. In this paper, when constructing the graph, we adopted the Model I in PGNN. Meanwhile, graph assumes that the chip has multi-layer bidirectional wiring, where the first layer is for vertical wiring and the second layer is for horizontal wiring, and the modeling is carried out based on this assumption.

As shown in Fig. 7, in a grid $g(x,y)$, assume there is a set of pin points $P = \{P_1, P_2, \cdots P_i\}$. Each pin point represents a vertex in the grid $G_{(x,y)}(V, E)$. The information contained in the vertex corresponds to the physical information of the pin point. The following are the detailed details of $\vec{v_i}$.

1) **The $x$ or $y$ coordinate of the pin point.** The first value of the vertex is either the $x$-coordinate or the $y$-coordinate of the pin point. When the wiring method is horizontal wiring, it is the $x$-coordinate; when the wiring method is vertical wiring, it is the $y$-coordinate. Meanwhile, the coordinates are normalized to facilitate GNN convolution.

2) **The layer where the pin is located.** Since the pixel-based graph is applicable to multi-layer chips, the layer number where the pin is located is represented by a numerical value. At the same time, to improve the adaptability of the model, the layer number is also normalized.

3) **The density of obstacles in the adjacent two layers.** Let $A(v_i)$ represent the proportion of the total area of the pin point $Pi$ that is covered by the upper and lower layers. Given that the wiring methods of the upper and lower layers are different, the situations of both layers are considered, and the average value of the obstacle densities of the two layers is taken. For example, as shown in Fig. 7, $P_{11}$ represents the pin point on the first layer, $P_{21}$ represents the pin point on the second layer, and $P_{31}$ represents the pin point on the third layer. $A(P_{21})$ is calculated by taking the proportion of the area of the pin points on the first and third layers that cover the pin point on the second layer, and then dividing this value by 2.

The edge $e_{ij}$ represents an edge connecting nodes $v_i$ and $v_j$. Nodes $v_i$ and $v_j$ are connected only when the distance between them is less than 1 μm and they belong to the same layer. When the distance between $v_i$ and $v_j$ is less than 1 μm, it indicates that the influence between the pin points strongly affects the connectivity, and the strength of this influence is described by the attributes of the edge. The following is the specific edge-modeling method.

1) **The specific distance between two nodes.** The physical distance between nodes can intuitively reflect the degree of influence between them. The closer the distance, the greater the degree of influence. At the same time, for the convenience of GNN calculations, we have normalized the distance.

2) **The relative position $\vec{d_{ij}}$ between two connected pin points.** We follow the method proposed in PGNN. In this method, FLUTE [20] and edge shifting techniques [21] are used to quickly find the correct way to access the pins. Different from PGNN, in pixel-based graph, only two modes are defined to represent whether it is the correct direction. For example, in Fig. 7, if the correct direction of $p_{12}$ is upward, then is 1; if the correct direction of $P_{12}$ is downward, then $\vec{d_{12}}$ is 0.

Due to the complexity of chip layouts, the number and size of graphs vary across different designs. To handle this, we design a mapping that allows a single GNN to process graphs of varying shapes and quantities.

This spatial guidance map is generated alongside the pixel-based graph. The map encodes the spatial location of each tile and provides a means to align GNN outputs back to the global layout. For tiles without any pins, empty graphs are generated to maintain spatial consistency.

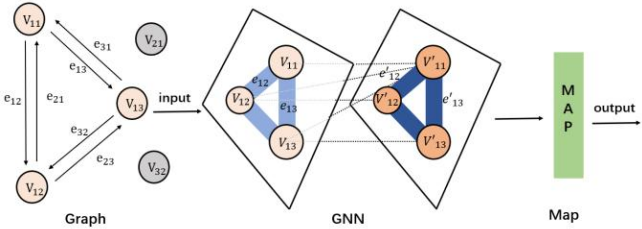

Fig. 8 The work of pixel-based graph & Supporting GNN

As shown in Fig. 8, In GNN, each edge $e_{ij}$ between nodes $v_i$ and $v_j$ carries information about obstacles or interference when routing from $v_j$ to $v_i$. A learned function $f_{cdot}$ is used to compute the edge embedding:

$$V_{ij} = f(V_i \ || \ V_j \ || \ e_{ij}) \quad (5)$$

The aggregated influence from all neighboring nodes is then computed as:

$$e_i = \sum_{j=1}^{n} V_{ij} \quad (6)$$

Finally, node $v_i$ is updated using another learnable function $f'$:

$$V'_i = f'(e_i \ || \ V_i) \quad (7)$$

This message-passing mechanism enables GNN to capture both local pin interactions and global topological constraints. After three layers of graph convolution, each pixel-based graph produces a learned representation which is projected back into a 2D layout grid using the map.

*C. Map-Guided Integration of U-Net and GNN Features*

In multi-modal neural network architectures, combining information from heterogeneous sources—such as image-based features and graph-based representations—has become an effective strategy for improving model performance in structured prediction tasks [6, 23]. Feature fusion refers to the process of integrating these diverse representations into a unified latent space, enabling downstream layers to make predictions based on both modalities.

To enhance the interaction between branches, fusion can be attention-guided. In such cases, one branch (e.g., graph-based reasoning) can generate importance scores or attention maps that modulate the feature representations in another branch (e.g., convolutional networks). This process enables one modality to inform or refine the feature weighting of another, improving interpretability and focus.

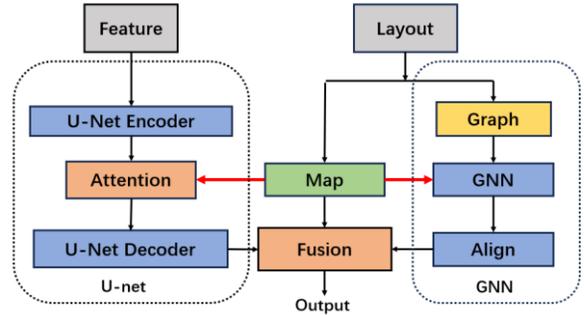

Fig. 9 The main position of map.

Once aligned, the fused features are typically passed through a prediction head—often a shallow convolutional or fully connected network—that learns to jointly interpret the combined semantic and structural information as shown in Fig. 9. This interaction mechanism enables the model to leverage both local texture information and global relational cues, which is particularly useful in tasks involving spatial constraints, connectivity patterns, or rule-based violation detection.

The map plays a critical role in aligning the outputs of MD-Unet and tile-level GNN. First, it enables the spatial alignment of GNN outputs with the MD-Unet feature map. More importantly, it also guides the attention mechanism in MD-Unet, thereby enabling cross-modal feature enhancement.

Specifically, the channel attention mechanism is modified as:
$$M_c(F, S) = MLP\big(AvgPool(F) + W_s \cdot AvgPool(s)\big) \quad (8)$$

Here, F represents the MD-Unet feature map, S is the map, and $W_s$ is a learnable parameter. MLP denotes a two-layer fully connected network.

Similarly, the spatial attention is guided by:
$$M_s(F, S) = f([AvgPool(F); S]) \quad (9)$$

Formula (9) shows that the Map guides the spatial attention mechanism. This is achieved by directly concatenating the Map into the input feature $F$.

## D. Output Fusion and Final Prediction

After extracting spatial and structural features separately through MD-Unet and GNN, the final stage of MAGNet involves fusing their outputs to produce an accurate DRC violation prediction map. This fusion is critical, as it enables the model to simultaneously leverage the semantic-level spatial information learned by MD-Unet and the graph-level topological insights captured by GNN.

- **Output Alignment and Concatenation**

To ensure compatibility, the outputs of MD-Unet and GNN are first aligned to the same spatial resolution: The output of MD-Unet is a dense feature map of shape [256×256×1], representing the pixel-wise probability of DRC violations. The output of GNN is a sparse structural prediction map. As GNN operates on non-overlapping tiles, its output is first projected back into a [256×256] grid using the map, which stores the spatial positions of each tile.

Once both outputs are spatially aligned, they are concatenated channel-wise to form a fused tensor of shape [256×256×2].

- **Discriminator Head**

The fused feature tensor is passed to a lightweight discriminator head, which acts as the final decision layer. This head consists of the following components: A [3×3] convolutional layer with 16 filters and ReLU activation, designed to model the interaction between MD-Unet and GNN features. A [1×1] convolutional layer with 1 output channel, reducing the output back to a scalar field of shape [256×256×1]. A sigmoid activation function, mapping the output to the probability range [0, 1], indicating the likelihood of each pixel being a hotspot.

The sigmoid output represents a soft prediction map. To obtain the final binary hotspot map, a fixed threshold of 0.1 is applied:

$$\begin{cases} 1 & if\, P(x,y) \geq 0.1 \\ 0 & otherwise \end{cases}$$

This relatively low threshold is chosen to ensure high recall, minimizing the risk of missing potential violations, which is often preferred in layout validation.

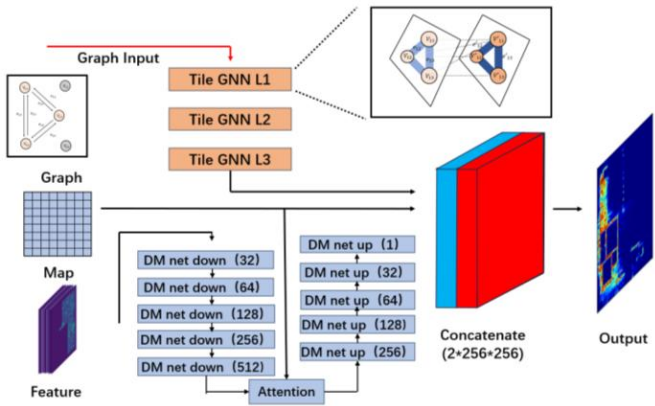

Fig. 10 The overall Structure of MAGNet.

- **Visual Illustration**

As shown in Fig. 10, the fused architecture processes each layout sample through parallel MD-Unet and GNN branches, followed by output fusion and a joint classifier. The resulting hotspot map closely aligns with ground truth.

## IV. EXPERIMENTAL VALIDATION

### A. Model Training Strategy

The training of MAGNet is conducted in two stages:

- **Stage 1: Pretraining MD-Unet:**

In the first stage, we independently train the modified U-Net (MD-Unet) to learn spatial features from the layout data. Training is conducted on an NVIDIA GeForce RTX 3070Ti GPU with a batch size of 16 and 512 iterations per epoch, totaling 510 epochs.

A three-phase learning schedule is applied:

*Epochs 1–10:* To avoid convergence to trivial solutions caused by sparse violation labels, all label values are amplified by a factor of 10 to ensure meaningful gradient flow.

*Epochs 11–200:* Original label values are restored, and the learning rate is set to $10^{-6}$.

*Epochs 201–510:* The learning rate is decreased to $10^{-7}$ for fine-tuning until convergence.

This staged training stabilizes MD-Unet's learning and ensures that it forms a robust feature extractor for spatial patterns.

- **Stage 2: Joint Training with GNN**

In the second stage, we integrate the GNN branch and train the full MAGNet in a joint learning setting. Due to the variable number and size of graphs per layout, the batch size is set to 1 to accommodate graph heterogeneity.

The training proceeds in two phases:

*Initial Phase:* The pretrained MD-Unet weights are frozen, allowing GNN to adapt to the learned spatial representations without disrupting them.

*Fine-tuning Phase:* Both MD-Unet and GNN are trained jointly to optimize cross-modal feature fusion.

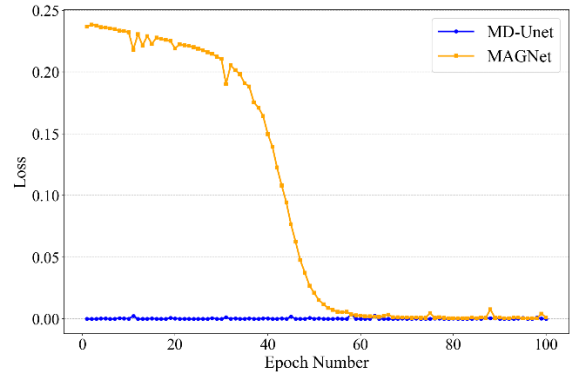

Fig. 11 The loss when training MD-Unet and MAGNet.

Unlike the first stage, label amplification is no longer required, as the MD-Unet is already well-initialized. As shown in Fig. 11, MAGNet achieves significantly lower training

loss—approximately one order of magnitude less than MD-Unet alone—demonstrating the effectiveness of incorporating topological features through GNN.

### B. Experimental Setup

The feature map extracted by MAGNet is the number of DRC violations, which is a continuous variable. To verify the accuracy of this continuous variable, we use Normalized Root Mean Square Error (NRMSE) and Structural Similarity Index Measure (SSIM) as evaluation metrics [7, 8]. Meanwhile, the goal of DRC is to identify hotspots, and hotspots are discrete variables. Referring to the settings in the reference papers, we set 0.1 as the threshold and use metrics such as True Positive Rate (TPF), F1, Precision, and False Positive Rate (FPR) for judgment.

*1) NRMSE*

NRMSE is a statistical indicator used to measure the degree of difference between predicted values and true values, and it is widely applied in many fields. Its specific calculation method is as the (9).

$$NRMSE = \frac{1}{y_{max}-y_{min}} \sqrt[2]{\frac{\sum_{i=0}^{W}\sum_{j=0}^{H}(y_{i,j}-\overline{y_{i,j}})}{H*W}} \quad (9)$$

The final NRMSE is obtained by calculating the NRMSE of each predicted map and the corresponding actual map, and then taking the average. The NRMSE can most directly reflect the fitting ability of the model to the labels. A smaller NRMSE indicates a smaller error during fitting. Therefore, it is of great significance to refer to this index when dealing with continuous quantities.

*2) SSIM*

SSIM is an index used to measure the similarity between two images and is widely applied in fields such as image quality assessment. It evaluates the degree of similarity between images by comparing three aspects: luminance, contrast, and structure. The specific calculation formula is as (10).

$$SSIM = \frac{(2\mu_x\mu_y+C_1)(2\sigma_{xy}+C_2)}{(\mu_x^2+\mu_y^2+C_1)(\sigma_x^2+\sigma_y^2+C_1)} \quad (10)$$

The calculation of SSIM is relatively complex and involves the values of Win_Size and Data_Range. In this paper, Win_Size is set to [11×11] and Data_Range is set to 1. The SSIM takes into account the degree of similarity of the overall structure of an image. Images with a high SSIM have an advantage when it comes to direct visual judgment by the human eye. Therefore, the SSIM is also a meaningful evaluation index for models.

*3) TPR, F1, Precision, and FPR*

TPR, F1, Precision, and FPR are common metrics for evaluating the performance of classification models. They are used to assess the performance of models in different aspects and are mostly applied to binary classification problems. Their specific calculation methods are as (11).

$$TPR = \frac{TP}{TP+FN}$$

$$FPR = \frac{FP}{FP+TN}$$

$$\text{Precision} = \frac{TP}{TP+FP} \quad (11)$$

$$F1 = 2*\frac{\text{Precision}*\text{Recall}}{\text{Precision}+\text{Recall}}$$

TPR, F1 score, Precision, and FPR are all standard metrics for evaluating the detection accuracy of binary variables. When conducting DRC inspections, it is preferable to have more detections rather than missing any issues. Hence, a model with a higher TPR value is considered better. Conversely, a relatively high FPR may lead to false positives, so it is necessary to control the magnitude of the FPR. The F1 score and Precision can also serve as references to assess the predictive ability of the model.

### C. Experimental Results

The evaluation is conducted on the CircuitNet dataset, which provides pixel-level annotated features for DRC violation detection. In this comparative analysis, RouteNet [3], ibUnet [8], and Enhanced U-Net [7] are benchmarked against the proposed MD-Unet and MAGnet. The results are presented in Tables I and II. Since U-Net-based methods formulate the DRC problem as an image processing task, their outputs are continuous indicators. Consequently, Enhanced U-Net and ibUnet are evaluated solely using continuous metrics.

TABLE I COMPARISON OF CONTINUOUS INDICATORS

| Models | Avg NRMSE(%) | Avg SSIM |
|---|---|---|
| RouteNet | 3.82 | 97.11 |
| ibUnet | 3.04 | 97.62 |
| Enhanced U-Net | 2.90 | 97.72 |
| MD-Unet | 2.86 | **98.11** |
| MAGnet | **1.75** | 97.93 |

TABLE II COMPARISON OF DISCRETE INDICATORS

| Model | FPR | TPR | Accuracy | AUC | F1 | Precision |
|---|---|---|---|---|---|---|
| RoutNet | 4.4 | 82.1 | 95.8 | 0.93 | 55.70 | 59.16 |
| J-net | 5.0 | 49.6 | 86.5 | 0.84 | 57.48 | 61.67 |
| MD-Unet | 10.34 | **99.53** | **99.22** | 0.83 | 64.23 | 81.66 |
| MAGnet | **3.5** | 98.11 | 98.89 | 0.82 | **69.72** | **91.00** |

When comparing the continuous indicators with ibUnet and RouteNet, the NRMSE decreased by 1.3%. The specific experimental results are shown in the table. When comparing the discrete indicators with MD-Unet and J-NET, it can be seen that for MD-Unet, the TPR, Accuracy, F1, and Precision have relatively significant improvements. In particular, the TPR is over 99%, indicating that almost all hotspots have been identified. However, the FPR has an undesirable increase, suggesting that the model is somewhat overly sensitive and misidentifies some normal points. We can see the certain result in Table I and Table II.

Compared with MD-Unet, MAGnet has a lower FPR, which is reduced by approximately 50%. At the same time, the TPR of MAGnet only decreases by less than 1%. After adding GNN, the F1 index and Precision of the model have also been further improved. This indicates that GNN effectively guides the attention mechanism of the U-Net, enabling the U-Net to pay more attention to important indicators and successfully reducing the FPR.

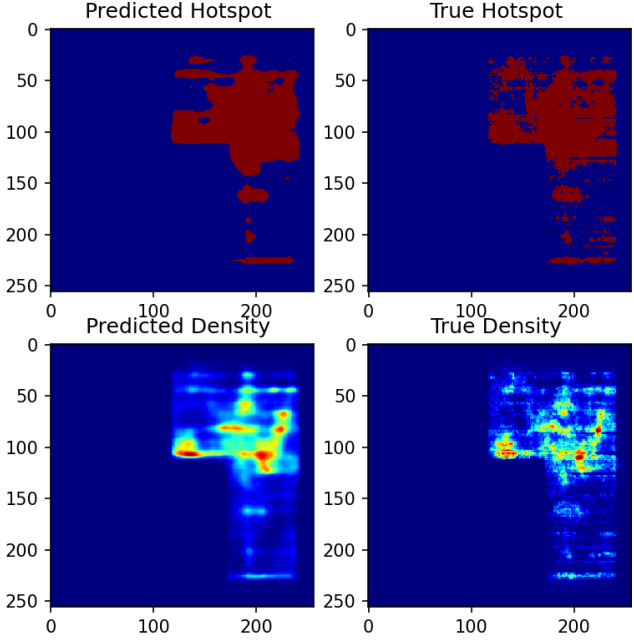

Fig. 12 The final effect of MAGNet.

We can see the final effect in Fig. 12, The two figures above are the actual prediction maps of hotspots, and the two figures below are the actual prediction maps of density. It can be seen that the hotspots predicted by MAGnet are more continuous and complete regions. However, there are many small areas in the accurate values of the hotspots, which are difficult for the model to predict. These areas are aspects that can be further improved in subsequent research.

*D. Experimental Analysis*

As shown in the Table II, the change in continuous indicators (such as accuracy, precision, and F1-score) of MD-Unet after incorporating the dynamic attention and multi-scale convolution modules is relatively small. This suggests that while the improvements in general performance are incremental, the model becomes more robust in handling complex patterns. However, due to the introduction of the multi-scale convolution mechanism, which expands the receptive field and captures features at different spatial scales, the model size and computational complexity have increased correspondingly.

Despite this, MD-Unet exhibits excellent sensitivity to hotspots, accurately identifying almost all true positive regions. We attribute this behavior primarily to the label amplification strategy applied during the early phase of training—specifically, the multiplication of DRC labels by a factor of 100. This operation effectively strengthened the learning signal associated with minority positive samples (i.e., the sparse DRC violations), enabling the model to better capture low-frequency but critical features. Consequently, the model develops a stronger internal representation of hotspot patterns. Meanwhile, the increase in FPR resulting from label amplification remains within an acceptable range, possibly due to the incremental training strategy that helped stabilize learning and prevent overfitting.

After retraining with the GNN module integrated, the model's ability to distinguish between true and false DRC violations further improved. Specifically, the misjudgment rate of non-hotspot regions was significantly reduced. This enhancement is largely due to the introduction of *pixel-based graph*, which explicitly encodes spatial connectivity, metal layer distribution, and pin-level features—structural information that is complementary to pixel-level visual cues captured by the U-Net. By guiding the U-Net's attention mechanism through graph-derived embeddings, the proposed MAGNet is able to focus on electrically and structurally critical areas, leading to more accurate predictions.

## V. CONCLUSIONS

In this paper, we propose MAGNet, a hybrid architecture that integrates U-Net with a tile-level GNN module for DRC hotspot prediction. By combining spatially aligned feature maps with graph-based structural priors, MAGNet achieves superior performance in both continuous and discrete prediction tasks. However, MAGNet also has certain limitations: the use of multi-scale convolution and graph modules increases model complexity and training time, and the variability in graph sizes limits training efficiency due to irregular batch processing. Moreover, the incremental training strategy may amplify noise under some configurations, as observed in the increased FPR of MD-Unet.

Through techniques such as label amplification, incremental training, and graph-guided attention, MAGNet outperforms baseline models including ibUnet, RouteNet, and MD-Unet. On continuous metrics, MAGNet achieves the lowest NRMSE of 1.75%, indicating a stronger regression capability in learning DRC density distributions. On discrete metrics, MAGNet maintains a high TPR of 98.11%, while effectively reducing the FPR to 3.5%, which is approximately 50% lower than that of MD-Unet. Furthermore, MAGNet records the highest F1 score and precision among all models, suggesting its robustness in hotspot classification and reduced false detections.

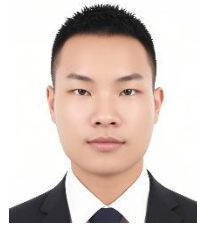

**WeiHan Lu** was born in China in 2004. He is currently an undergraduate student majoring in automation at the School of Automation, Southeast University, Nanjing, China, and is expected to graduate in 2026.

His current research interests primarily focus on neural networks and EDA, where he aims to explore innovative applications and design methodologies.